\documentclass{IEEEtran}
\usepackage{cite}
\usepackage{graphicx}
\usepackage{mdwmath}
\usepackage{amssymb}
\usepackage{mdwtab}
\usepackage{stfloats}
\usepackage[tight,footnotesize]{subfigure}
\usepackage{amsmath,amsthm}
\usepackage{url}

\begin{document}

\title{\LARGE Fundamental Wireless Performance of a Building}
\author{Jiliang~Zhang,~\textit{Senior Member~IEEE},~Andr\'es~Alay\'on~Glazunov,~\textit{Senior Member IEEE}, \\ Wenfei Yang,~\textit{Graduate Student Member IEEE},
 and Jie~Zhang,~\textit{Senior Member IEEE}
 \vspace{-0.3in}\
 \thanks{The research is funded by EUROSTARS Project Build-Wise (11088). Corresponding Author: Prof. Jie Zhang}
\thanks{
Jiliang Zhang and Wenfei Yang are  with the Department of Electronic and Electrical Engineering, The University of Sheffield, Sheffield, UK. 

Andr\'es Alay\'on Glazunov is with Department of Electrical Engineering, University of Twente, Enschede, Overijssel, Netherlands, and also with Department of Electrical Engineering, Chalmers University of Technology, Gothenburg, Sweden.

Jie Zhang is with the Department of Electronic and Electrical Engineering, the University of Sheffield, Sheffield, UK, and also with Ranplan Wireless Network Design Ltd., Cambridge, UK.

}
}
\markboth{Accepted at IEEE Wireless Communications}
{Shell \MakeLowercase{\textit{et al.}}: Bare Demo of IEEEtran.cls for Journals}
\maketitle
\begin{abstract}
Over 80\% of wireless traffic already takes place in buildings.
Like water, gas, and electricity, wireless communication is becoming one of the most fundamental utilities of a building.
It is well known that building structures have a significant impact on in-building wireless networks. 
If we seek to achieve the optimal network performance indoors, the buildings should be designed with the objective of maximizing wireless performance. 
So far, wireless performance has not yet been considered when designing a building. 
In this paper, we introduce a novel and interdisciplinary concept of building wireless performance (BWP) to a wide audience in both wireless communications and building design, emphasizing its broad impacts on wireless network development and deployment, and on building layout/material design. We first give an overview of the BWP evaluation framework proposed in our state-of-the-art works and explain their interconnections. Then, we outline the potential research directions in this exciting research area to encourage further interdisciplinary research.
\end{abstract}

\section{Introduction}
A recent survey shows that 80-96\% of mobile traffic takes place in buildings.
In the last few decades, the wireless communication industry has employed various strategies to provide in-building wireless coverage and capacity cost-effectively. 
In the 2G/3G eras, indoor coverage mainly relied on outdoor radio base stations (BSs).
Moving towards the 4G era, indoor networks become more widely deployed to compliment the outdoor-to-indoor coverage. 
Now, in the 5G era and beyond,  the ever boosting indoor coverage and capacity demands are expected to be satisfied by densified indoor BSs \cite{5Gdense1}. 

Currently, in-building wireless network design has been independent of building design and construction practice. 
However, essential factors considered by architects and civil engineers when designing the building structures, including building materials and building layouts, also inevitably affect the in-building network deployment and signal propagation, and hence the indoor network performance. 
The potential limits imposed by building structures on indoor wireless networks can only be overcome by modifying the building itself.
Therefore, in order to quantify the effects of building structures on the indoor network performance, or how wireless friendly a building is in an easy-to-understand term, we have introduced essential figures of merit to evaluate the building wireless performance (BWP)  \cite{BWP,IG,multiwall}.
The BWP should be considered at the design stage of a building, just as the structural safety and energy efficiency performance. 
The urgency and timeliness of the introduction of the BWP is discussed next.  

Firstly, as the indoor usage of wireless services increases, it will be no longer feasible to meet indoor wireless traffic demands by exclusively optimizing the in-building wireless network deployment cost-effectively \cite{6G2}.
The wireless network performance in an existing building may reach, at best, the upper bound constrained by the current building structures independent from the network optimization efforts.
Therefore, the design of wireless friendly buildings is indispensable to meet the ever boosting indoor traffic demands in the next decade and beyond.

Secondly, the energy efficiency of in-building wireless networks has become a matter of great concerns in both the wireless communication and building design industries. 
At present, mobile operators are already among the top energy consumers. 
The power consumption by an indoor 5G network is expected to be 10~$\mathrm{Watt/m^2}$~=~87.6~$\mathrm{kWh/m^2/year}$ as indicated in \cite[Table~1]{BuildingEnergy3}. 
Whereas, according to the average energy consumption of different types of buildings listed in \cite[Table~6]{BuildingEnergy1},  the average energy consumed by a dwelling building in the USA is 147~$\mathrm{kWh/m^2/year}$ in 2003.
Moreover, an increasing trend of power consumption of wireless networks can be observed.  
In the upcoming 6G era, energy efficiency of wireless networks is expected to become from 10-100 times that of 5G \cite{6G2}. 
To achieve this goal in indoor built environments, the building structures shall be as wireless efficient as possible. Therefore, the time has come for the wireless friendly buildings!

Some related works do exist in the wireless communication sector investigating the impact of the environment on the indoor network performance; for example, fitting shadowing effects to a function of a random variable, characterizing indoor radio propagation channels with site-deterministic approaches, and modeling the distribution of the blockages by a random process are presented in \cite{AAGbook, 4wall}. However, they can only be applied to existing buildings or hypothetical indoor propagation scenarios rather than actual building structures.

The main objective of the article is to give an overview of the novel BWP concept and the vision to realize its potential.

 \section{ Impact of building structures on intended signal power and interference power}

In the in-building scenario, the signals can experience various propagation mechanisms triggered by building structures. 
For example, when a direct unobstructed path exists between the transmit antenna and the targeted user equipment (UE), i.e., under a line-of-sight (LOS) propagation scenario, the received signal power may be enhanced due to reflections from the floor, the ceiling, and the walls, as well as the waveguide effect in corridor-like environments.
Experiments have shown that the path loss exponent (PLE) can be less than 2, which is the theoretical PLE in free space, in LOS indoor environments in both low frequencies and the millimeter-Wave (mm-Wave) bands \cite{AAGbook}.
When obstacles block the direct path, i.e., under a non-line-of-sight (NLOS) propagation scenario, the received signal power will be attenuated due to the penetration loss.  
The reflection loss and penetration loss vary with the thickness and electromagnetic (EM) properties of building materials.
The power enhancement or attenuation could happen for either intended signals or interference signals of a typical indoor user. 
The signal-to-interference-and-noise ratio (SINR) is affected and hence the network performance, too.

In this section, we discuss the BWP by quantifying the impact of building structures on the power levels of intended signals and interference in a reference UE location.

\subsection{BWP metric design}

We define the metrics of BWP evaluation based on the comparison of the network performance in the in-building scenario with that in a benchmark scenario--the open-space, where no obstruction exists.	
Two BWP metrics, the interference gain (IG) and the power gain (PG), are introduced by comparing the SINR between the in-building scenario and the open-space scenario \cite{BWP}.
The SINR of a typical UE in the in-building scenario is denoted by
$\gamma_\mathrm{B} = P_{\mathrm{B}}/{\left(I_{\mathrm{B}}+\sigma^2\right)}$,
where $P_{\mathrm{B}}$ and $I_{\mathrm{B}}$ denote the total received intended signal power and interference power in the in-building scenario, respectively, while $\sigma^2$ represents the thermal noise power.
The SINR for a typical UE in the open-space scenario is denoted by
$\gamma_\mathrm{O} = {P_{\mathrm{O}}}/{\left(I_{\mathrm{O}}+\sigma^2\right)}$, where $P_{\mathrm{O}}$ and $I_{\mathrm{O}}$ denote the total received intended signal power and the interference power in the open-space scenario, respectively. 
The PG and the IG are defined as follows. 
\begin{itemize}
	\item \textbf{Power gain:} The PG is defined as the ratio of the total intended signal power received in the in-building scenario to the total intended signal power received in open space, as $g_{\mathrm{P}}  \triangleq P_{\mathrm{B}}/{P_{\mathrm{O}}}$. 
	It quantifies how easily wireless signals can reach every part of a building.
	
	\item \textbf{Interference gain:}  The IG is defined as the ratio of the total interference power received in the in-building scenario and the noise power to the total interference power received in open space and the noise power, as $g_{\mathrm{I}}  \triangleq {\left(I_{\mathrm{O}}+\sigma^2\right)}/{\left(I_{\mathrm{B}}+\sigma^2\right)}$.
	It quantifies how much interference is blocked in a wireless network occupying a certain amount of spectrum and transmit power, which indicates the impact of a building on the energy efficiency of indoor wireless networks.
\end{itemize}
Then, the product $g_{\mathrm{P}}g_{\mathrm{I}}$ represents the effective change of the SINR at the reference UE location due to the presence of the building structures since the identity
$\gamma_\mathrm{B}= g_{\mathrm{P}}g_{\mathrm{I}}\gamma_\mathrm{O}$ exists.
The higher values of $g_{\mathrm{I}}$ and $g_{\mathrm{P}}$ are, the more positive effects the building structures show on the indoor wireless network performance.
The metrics $g_{\mathrm{I}}$ and $g_{\mathrm{P}}$ can be employed by the architects and the civil engineers in the building design process as the indicators to assess the design quality of a building in terms of the BWP.

\subsection{Network model for the intrinsic BWP evaluation}

\begin{figure*} [t]
	\centering
	\subfigure[]{\includegraphics [width=2.7in,trim=0 0 0 20,clip]{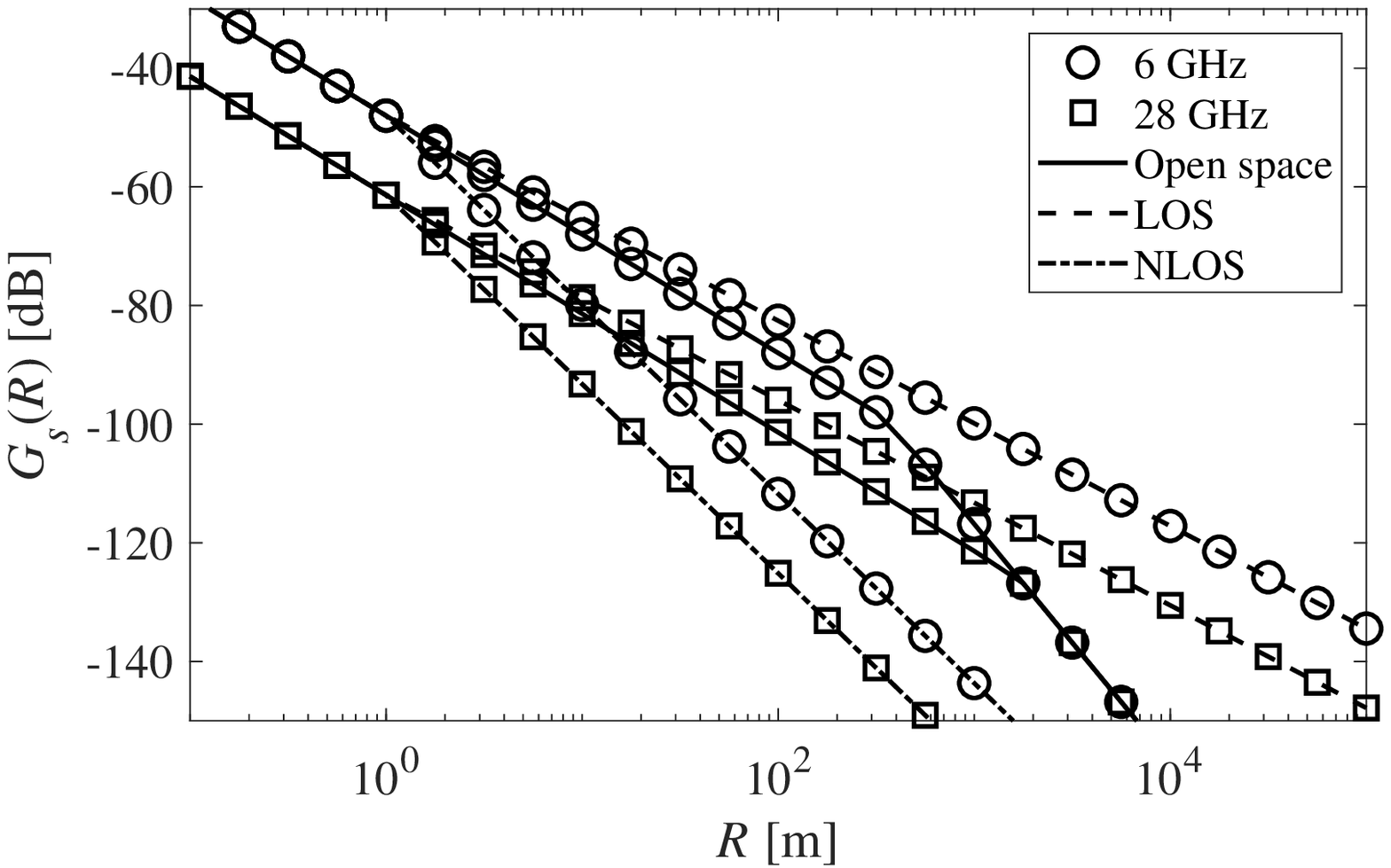}}
	\hspace{0.5in}
	\subfigure[]{\includegraphics [width=3 in,trim=0 0 0 10,clip]{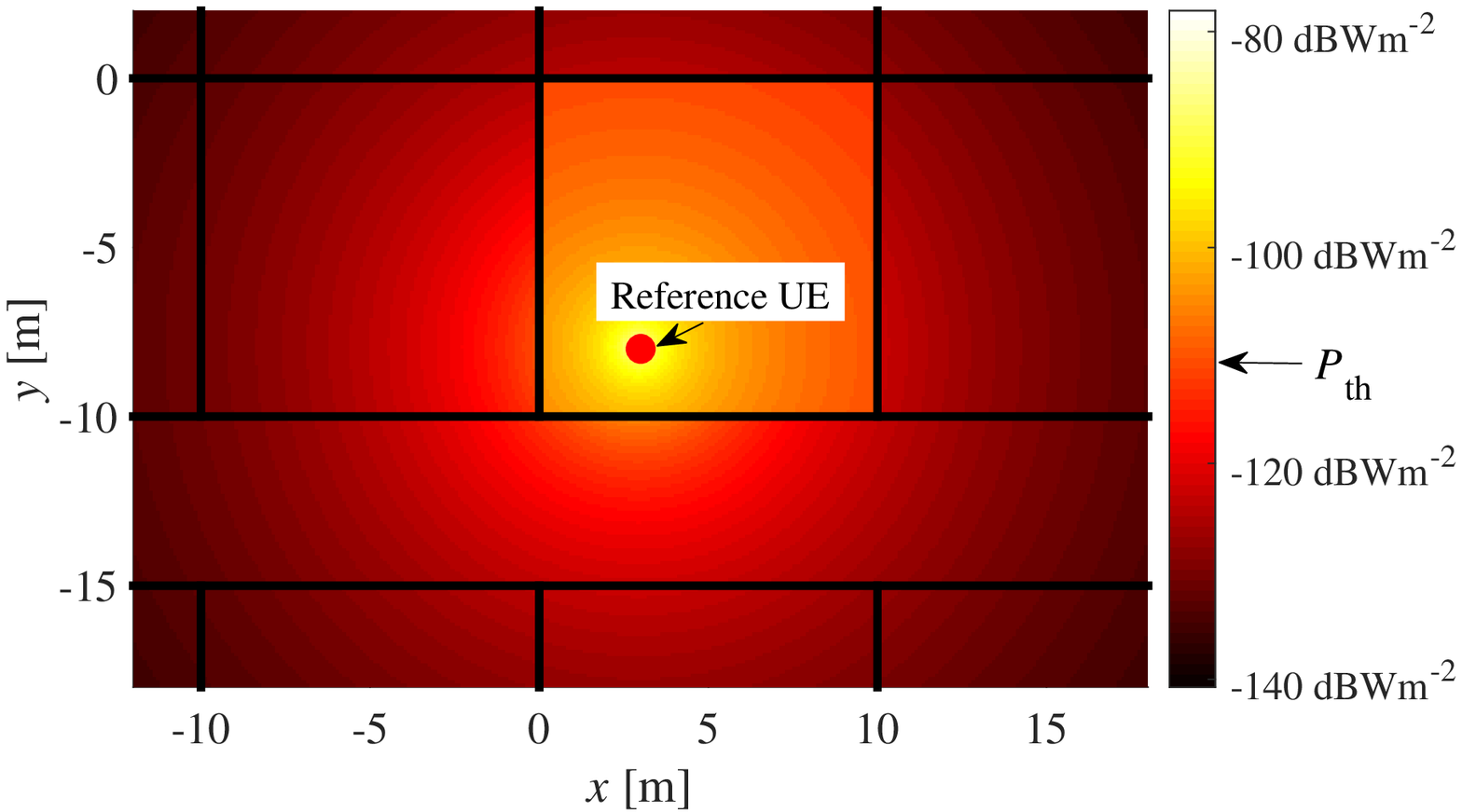}} \\	
    \caption{Model of extremely dense cooperative indoor network.
	(a) Path gain as a function of transmitter-receiver separation computed at 6~GHz and 28~GHz in the open-space, in-building LOS, and in-building NLOS scenarios. 
	(b) Receive power density on a given reference UE location at 28~GHz.
	} \label{model}
	\centering
	\subfigure[]{\includegraphics [width=2.2in,trim=0 0 0 0,clip]{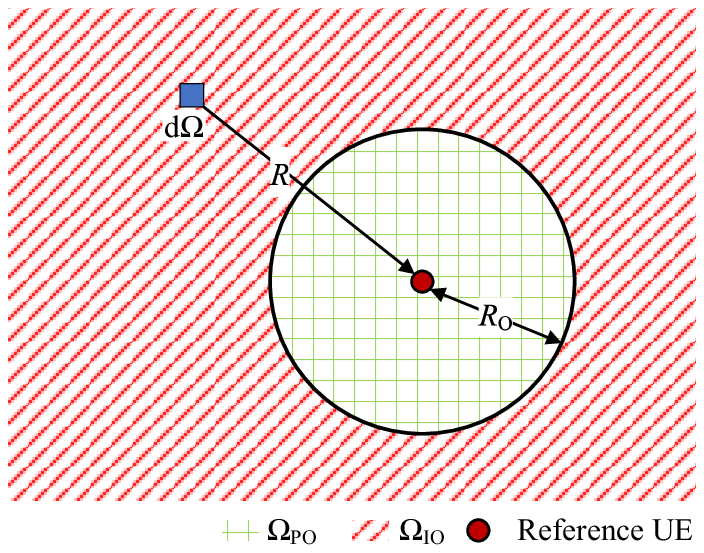}}
	\hspace{1in}
	\subfigure[]{\includegraphics [width=2.2in,trim=0 0 0 0,clip]{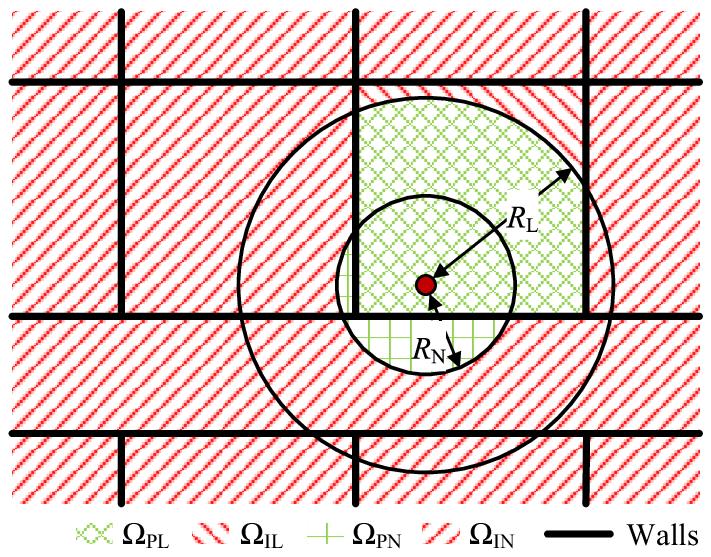}}\\
	\caption{ Examples of $\Omega_{\mathrm{PO}}$, $\Omega_{\mathrm{IO}}$, $\mathrm{\Omega}_{\mathrm{IL}}$,
		$\mathrm{\Omega}_{\mathrm{IN}}$,
		$\mathrm{\Omega}_{\mathrm{PL}}$, and
		$\mathrm{\Omega}_{\mathrm{PL}}$.
	(a) Open-space scenario. 
    (b) In-building scenario.} 
	\label{OpenVsBuilding}
\end{figure*}

There we focus on the effects of building structures on the maximum achievable performance gain of in-building wireless networks in the downlink transmissions. 
A general and idealistic network model has been employed to present the best case scenario of an indoor network taking into account fundamental network parameters.
Two main assumptions of the network model have been made as follows. 

Firstly, an infinite number of transmit elements are assumed in the network model.
The idea is to emulate small cells or, equivalently, single antennas, densely and uniformly distributed in a 2-D environment situated on the horizontal plane. 
Hence, in an arbitrary plane surface domain $\Omega$, the total transmit power from transmit elements within $\Omega$ is $P_{\mathrm{T}}A_\Omega$, where $A_\Omega$~$[\mathrm{m}^{2}]$ is the area of $\Omega$ and $P_{\mathrm{T}}$~$[\mathrm{W}\mathrm{m}^{-2}] $ denotes the transmit power density, i.e., the transmit power per unit area.
A remark is that this assumption can be considered  either from the perspective of distributed massive multiple-input and multiple-output (MIMO) systems \cite{Distributed2} or from the perspective of ultra-dense small-cell (UDS) networks \cite{5Gdense1}.
\begin{enumerate}
	\item \textbf{Distributed massive MIMO perspective:}
	Consider the downlink transmission in an indoor scenario. 
	Scaling up the number of antennas in an antenna array results in an increase of the physical size. In the limiting ideal case of an extremely massive MIMO scenario, an infinite number of antennas are uniformly distributed over the coverage area. 
	Hence, each pair of adjacent antennas has an infinitesimal space interval between them.
	For example, in an $A\times A$ square area, we deploy one distributed massive MIMO array with $N_{\mathrm{T}}$ antennas.  
	According to the power restriction, the transmit power of each antenna is $\frac{A^2P_{\mathrm{T}}}{N_{\mathrm{T}}}$, and each antenna occupies an area of $\frac{A^2}{N_{\mathrm{T}}}$. 
	When $N_{\mathrm{T}}$ goes to infinity, for an arbitrary infinitesimal area $\mathrm{d}\Omega$ in the scenario, the total transmit power is given by $P_{\mathrm{T}}\mathrm{d}\Omega$ as assumed above.
	
	\item \textbf{Ultra-dense small-cell perspective:}
	In the limiting case of the UDS scenario, an infinite number of small cells uniformly distributed over the coverage area is assumed. 
	Assuming that we deploy $N_{\mathrm{SC}}$ small cells in an $A\times A$ square area,  each small cell is allocated a transmit power of $\frac{A^2P_{\mathrm{T}}}{N_{\mathrm{SC}}}$. 
	In a small area $\mathrm{d}\Omega$ in the scenario, when $N_{\mathrm{SC}}$ goes to infinity, the total transmit power allocated to small cells in $\mathrm{d}\Omega$ is $P_{\mathrm{T}}\mathrm{d}{\Omega}$ on average. 
	And therefore, the ultra-densification of small cells leads to the assumed network model.
\end{enumerate}

Secondly, in a downlink transmission, the reference UE can make use of all the detectable power, which is constrained by the sensitivity of its receiver. 
The receive power is in turn assumed to be the result of maximum ratio transmissions (MRT), which maximizes the received signal-to-noise ratio (SNR) at the UE, in distributed massive MIMO networks or coordinated multipoint (CoMP) transmissions in UDS networks. 
In this article, the detectable power is defined as the power received from transmit elements satisfying the condition
$P_{\mathrm{T}}G_{s}(R)>P_{\mathrm{th}}$,
where $P_{\mathrm{T}}$ denotes, as above, the transmit density, $P_{\mathrm{th}}$~$[\mathrm{W}\mathrm{m}^{-2}] $ is a threshold power level determined by the sensitivity of the UE receiver, $R$~$[\mathrm{m}]$ is the link length from the transmit element to the UE, $G_{s}(R)$ is the path gain under the corresponding scenario $s\in\{\mathrm{O, L, N}\}$, denoting the open-space, in-building LOS, and in-building NLOS scenarios, respectively.  
In cooperative UDS or massive MIMO systems, channel responses are smoothed by the extremely large spatial diversity as a result of the favorable action of the law of large numbers. 
In essence, small-scale fading is negligible, therefore, only large-scale fading is considered.

In the open-space scenario, the received signal has two components, i.e., the LOS component and the ground reflection component. 
Thus, $G_\mathrm{O}(R)$ is computed using the two-ray ground-reflection model given by \cite[Eqs. (4,5)]{tworay}.
For the in-building LOS and NLOS scenarios, the multi-slope path gain models in \cite{Multisolpe} are employed to compute $G_\mathrm{L}(R)$ and $G_\mathrm{N}(R)$, where the numerical values of the PLEs $n_{\mathrm{L}}$ and $n_{\mathrm{N}}$ for LOS and NLOS scenarios, respectively, are determined by the characteristics of the propagation environments.
In this article, we specialize our results to $n_{\mathrm{L}}=1.73$, and $n_{\mathrm{N}}=3.19$ following the The 3rd Generation Partnership Project (3GPP) indoor channel model devised for a wide range of frequencies, i.e., 0.5-100~GHz \cite[Table 7.4.1-1]{3GPP}.
The numerical results are generated with 
$P_{\mathrm{T}}=-34\ \mathrm{dBWm^{-2}}$, $P_{\mathrm{th}}=-110\ \mathrm{dBWm^{-2}}$. The height of both transmit antennas and the receive antenna is set as $1.2$~m in both the 6-GHz and 28-GHz bands.
Path gains computed assuming the open-space, in-building LOS, and in-building NLOS scenarios at 6~GHz and 28~GHz are plotted in Fig.~\ref{model}(a).
Following the assumed network model, the receive power density in a given a reference UE location at 28~GHz is plotted in Fig.~\ref{model}(b).

\begin{figure*} [t]
	\centering
	\subfigure[]{\includegraphics[width=3.4in,trim=240 105 205 100,clip]{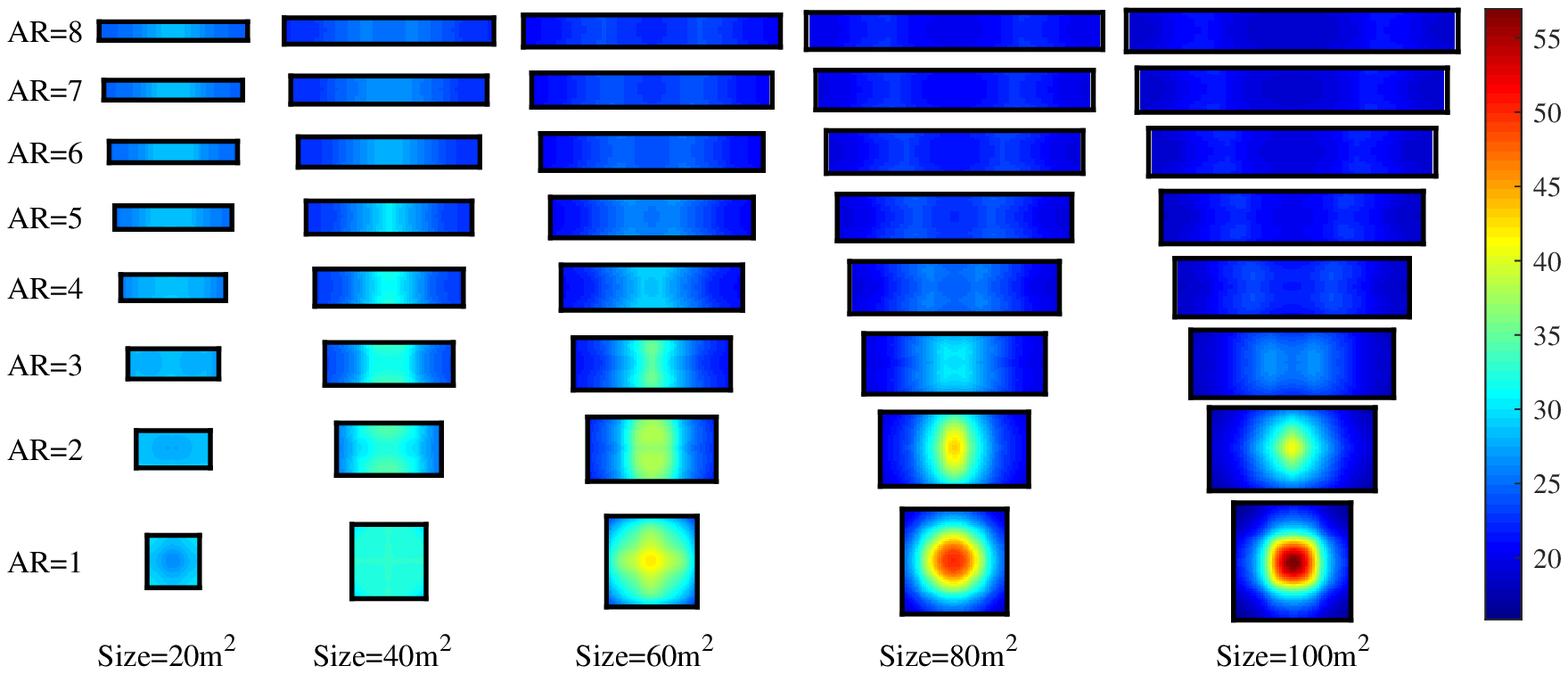}}
	\subfigure[]{\includegraphics[width=3.4in,trim=240 105 205 100,clip]{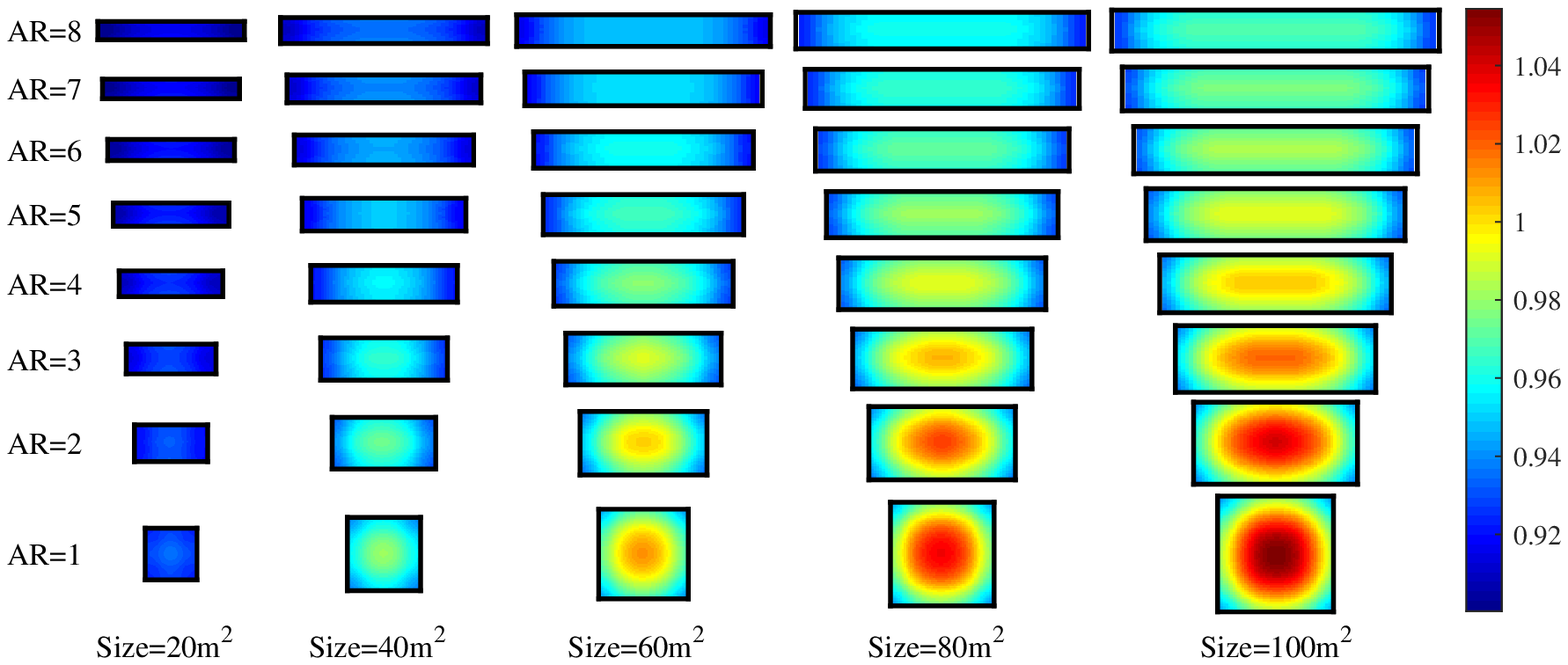}}\\
	\subfigure[]{\includegraphics[width=3.4in,trim=240 105 205 100,clip]{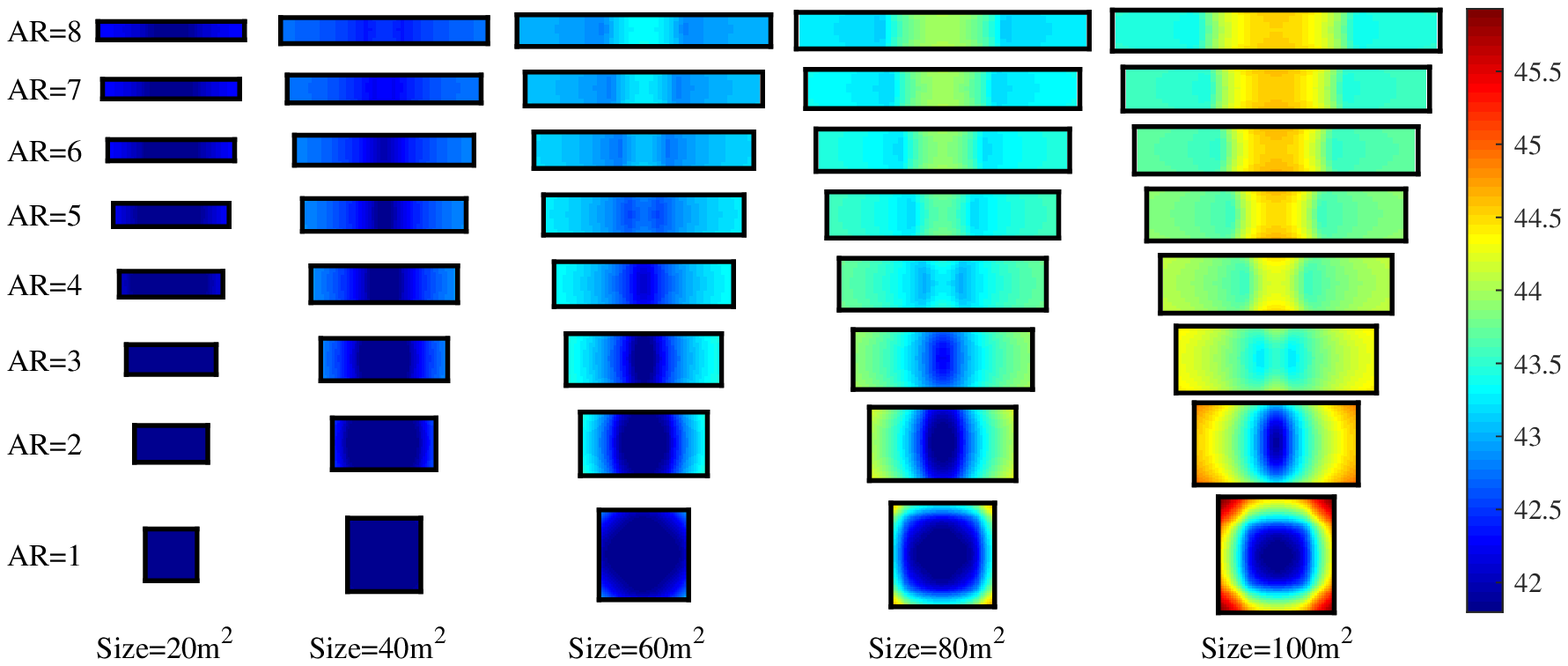}}
	\subfigure[]{\includegraphics[width=3.4in,trim=240 105 205 100,clip]{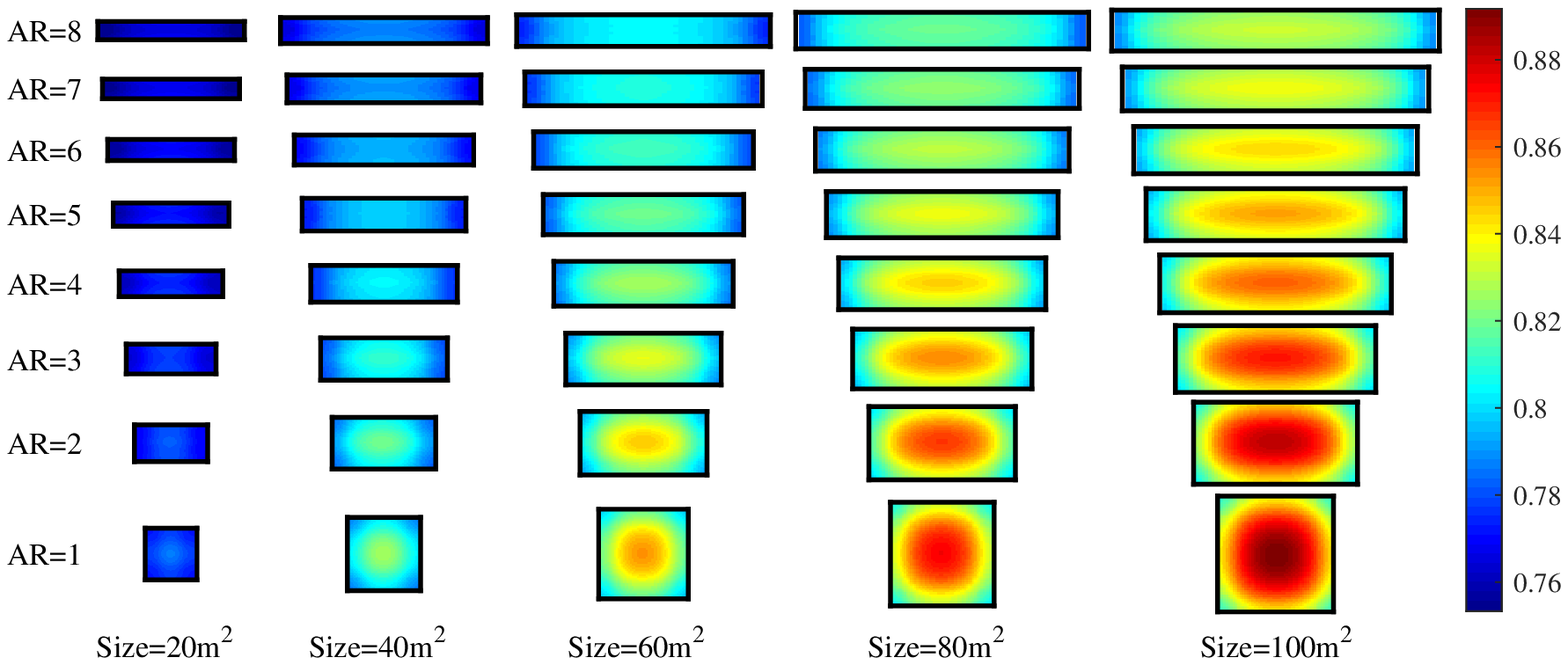}}\\	
	\subfigure[]{\includegraphics[width=1.65in,trim=0 0 0 20,clip]{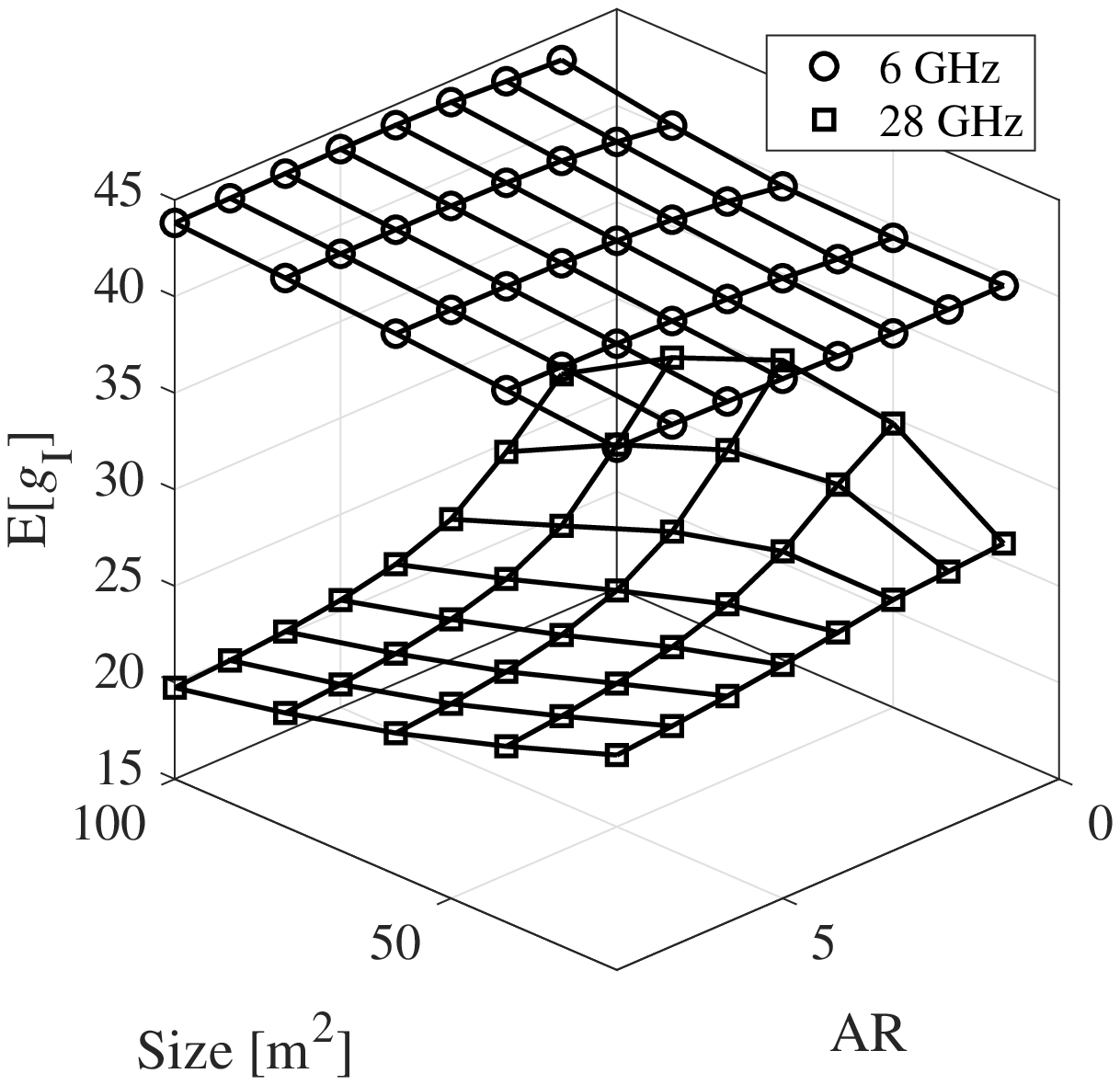}}
	\subfigure[]{\includegraphics[width=1.65in,trim=0 0 0 20,clip]{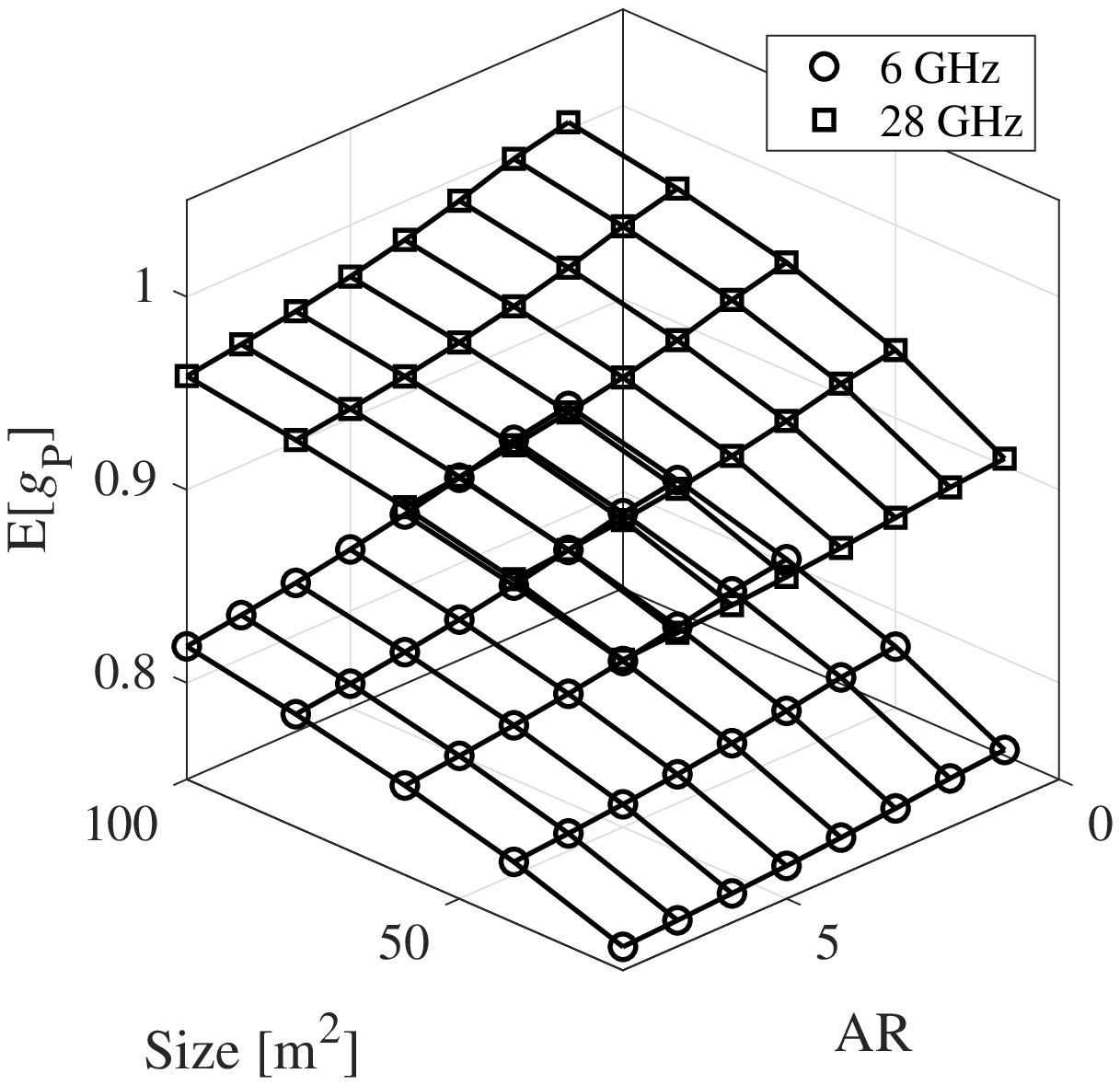}}
	\hspace{10pt}
	\subfigure[]{\includegraphics[width=1.6in,trim=0 0 0 20,clip]{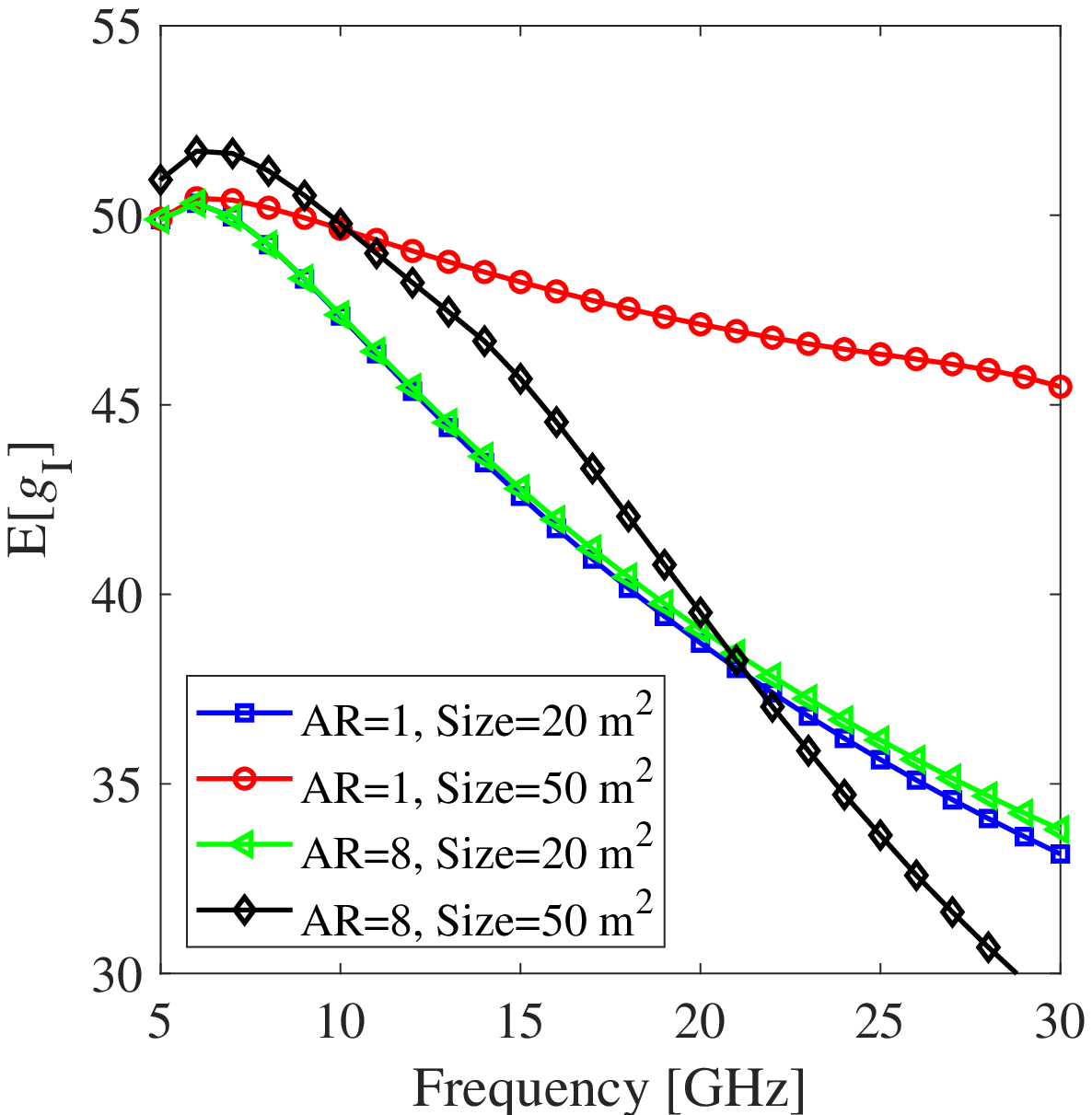}}
	\subfigure[]{\includegraphics[width=1.6in,trim=0 0 0 20,clip]{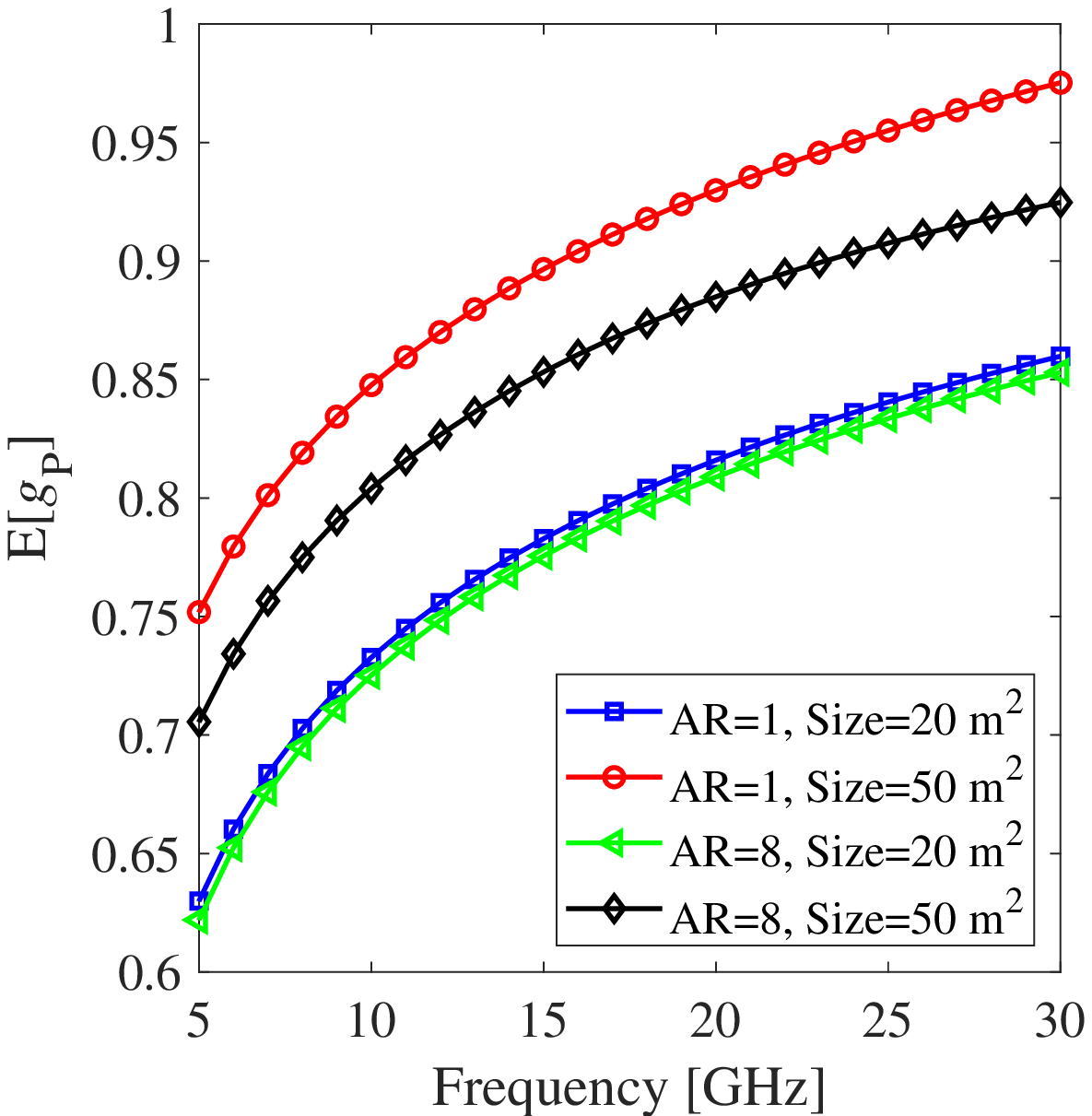}}\\
	\centering\caption{
        Evaluation of $g_{\mathrm{I}}$ and $g_{\mathrm{P}}$ in single rectangular rooms.
(a)~Distributions of $g_{\mathrm{I}}$ in single rectangular rooms at 28~GHz.
(b)~Distributions of $g_{\mathrm{P}}$ in single rectangular rooms at 28~GHz.
(c)~Distributions of $g_{\mathrm{I}}$ in single rectangular rooms at 6~GHz.
(d)~Distributions of $g_{\mathrm{P}}$ in single rectangular rooms at 6~GHz.
(e)~Average $g_{\mathrm{I}}$ as a function of the size and the AR.
(f)~Average $g_{\mathrm{P}}$ as a function of the size and the AR.
(g)~Average $g_{\mathrm{I}}$ as a function of the operating frequency.
(h)~Average $g_{\mathrm{P}}$ as a function of the operating frequency.
	} \label{gigp6G}
\end{figure*}

\subsection{Derivations of the BWP metrics}

We use $R_{\mathrm{O}}$, $R_{\mathrm{L}}$ and $R_{\mathrm{N}}$ to denote the coverage distances in the open-space, in-building LOS, and in-building NLOS scenarios, which are derived by solving $P_{\mathrm{T}}G_{s}(R_{s})=P_{\mathrm{th}}$.  
For an arbitrary UE location, the areas in open space containing the transmit elements sending intended and interference signals are denoted by $\Omega_{\mathrm{PO}}$ and $\Omega_{\mathrm{IO}}$, respectively, as the examples plotted in Fig.~\ref{OpenVsBuilding}(a). 
$P_{\mathrm{O}}$ and $I_{\mathrm{O}}$ can then be computed by the numerical evaluations of the integrals  $P_{\mathrm{O}}
=\int_{\Omega_{\mathrm{PO}}}
{P_{\mathrm{T}}}{G_{\mathrm{O}}(R)}\mathrm{d}\Omega$ and
$I_{\mathrm{O}}
=\int_{\Omega_{\mathrm{IO}}}
{P_{\mathrm{T}}}{G_{\mathrm{O}}(R)}\mathrm{d}\Omega$, respectively.
In the in-building scenario, signals reach the UE under LOS and NLOS propagation scenarios are referred to as the LOS signals and NLOS signals, respectively. 
Given a reference UE location in the in-building scenario, the areas sending LOS intended and interference signals are denoted by $\Omega_{\mathrm{PL}}$ and $\Omega_{\mathrm{IL}}$, respectively.
On the other hand, the areas sending NLOS intended and interference signals are denoted by $\Omega_{\mathrm{PN}}$ and $\Omega_{\mathrm{IN}}$, respectively. 
The examples of $\Omega_{\mathrm{PL}}$, $\Omega_{\mathrm{IL}}$, $\Omega_{\mathrm{PN}}$, and $\Omega_{\mathrm{IN}}$ for a given reference UE location are plotted in Fig.~\ref{OpenVsBuilding}(b). 
$P_{\mathrm{B}}$ and $I_{\mathrm{B}}$ can then be computed by the numerical evaluations of the integrals
$P_{\mathrm{B}}
=\int_{\Omega_{\mathrm{PL}}}
{P_{\mathrm{T}}}{G_{\mathrm{L}}(R)}\mathrm{d}\Omega +
\int_{\Omega_{\mathrm{PN}}}
{P_{\mathrm{T}}}{G_{\mathrm{N}}(R)}\mathrm{d}\Omega$ and
$I_{\mathrm{B}}
=\int_{\Omega_{\mathrm{IL}}}
{P_{\mathrm{T}}}{G_{\mathrm{L}}(R)}\mathrm{d}\Omega+
\int_{\Omega_{\mathrm{IN}}}
{P_{\mathrm{T}}}{G_{\mathrm{N}}(R)}\mathrm{d}\Omega$, respectively.
The closed-form expressions of $P_{\mathrm{O}}$, $I_{\mathrm{O}}$, $P_{\mathrm{B}}$, and $I_{\mathrm{B}}$ have been derived in \cite{BWP}, and then $g_\mathrm{I}$ and $g_\mathrm{P}$ can be computed by the definitions. 

Besides the analytical models, we have developed a machine-learning-based approach to enable fast BWP evaluation based on the building structures. 
Two two-hidden-layer fully-connected artificial neural networks (ANNs) have been trained for the 6-GHz and 28-GHz bands, respectively. 
The input layer consists of 4 units, including the coordinates of the UE and the dimensions of the room, each hidden layer consists of 30 units with rectified linear activation functions, and the outputs are the values of $g_\mathrm{I}$ and $g_\mathrm{P}$.

\begin{figure*}[t]
	\centering
	\subfigure[]{\includegraphics[width=4.6in,trim=120 90 0 100,clip]{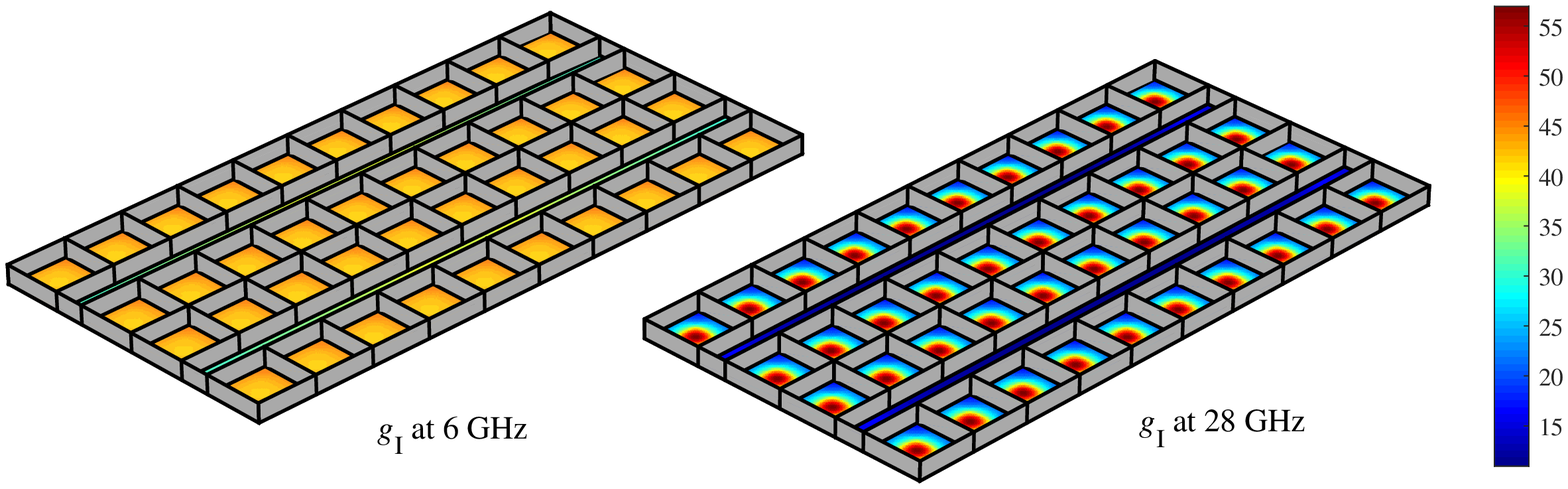}}
	\subfigure[]{\includegraphics[width=1.6in,trim=0 0 0 20,clip]{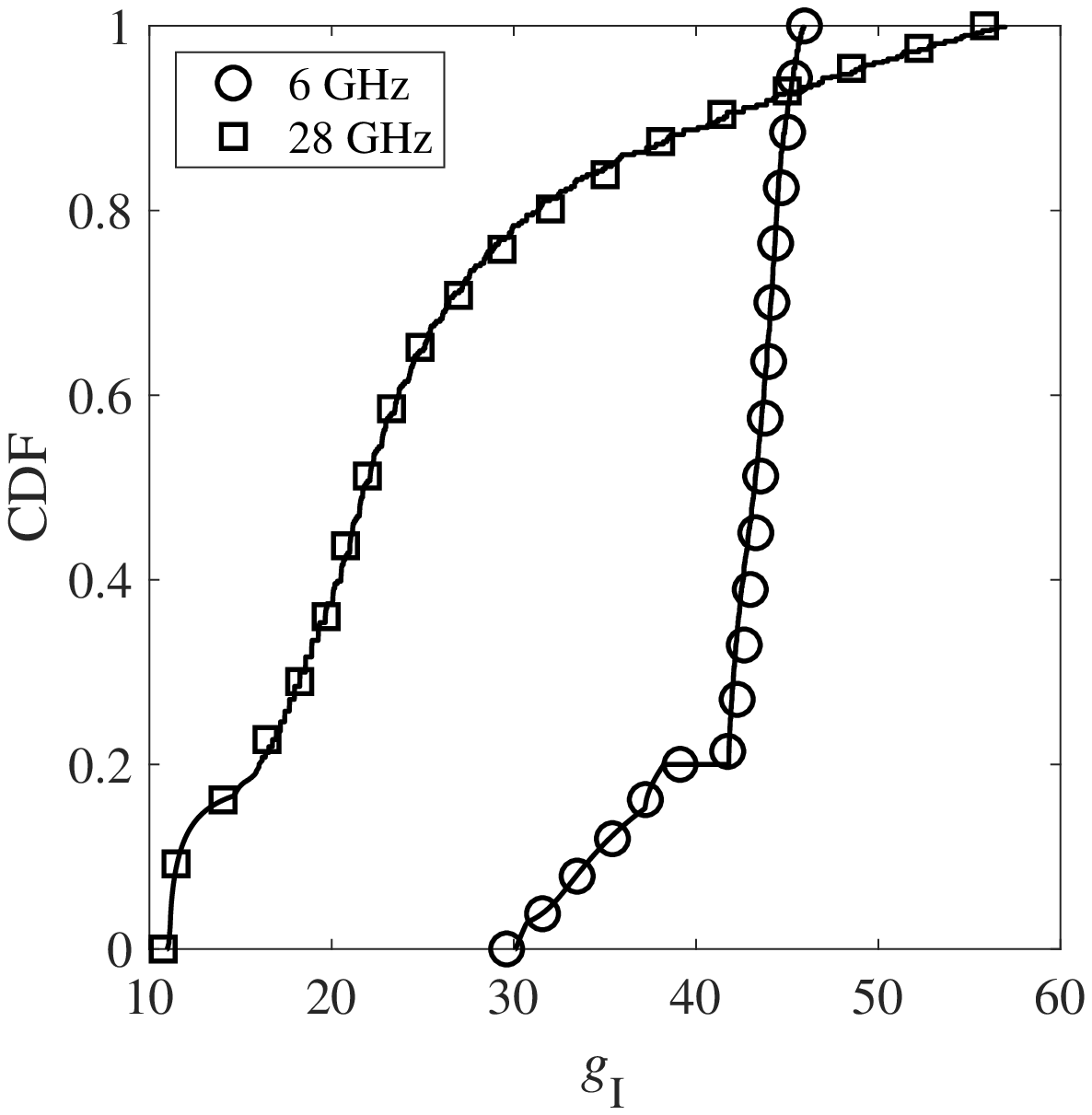}}	\\
	\subfigure[]{\includegraphics[width=4.6in,trim=120 90 0 100,clip]{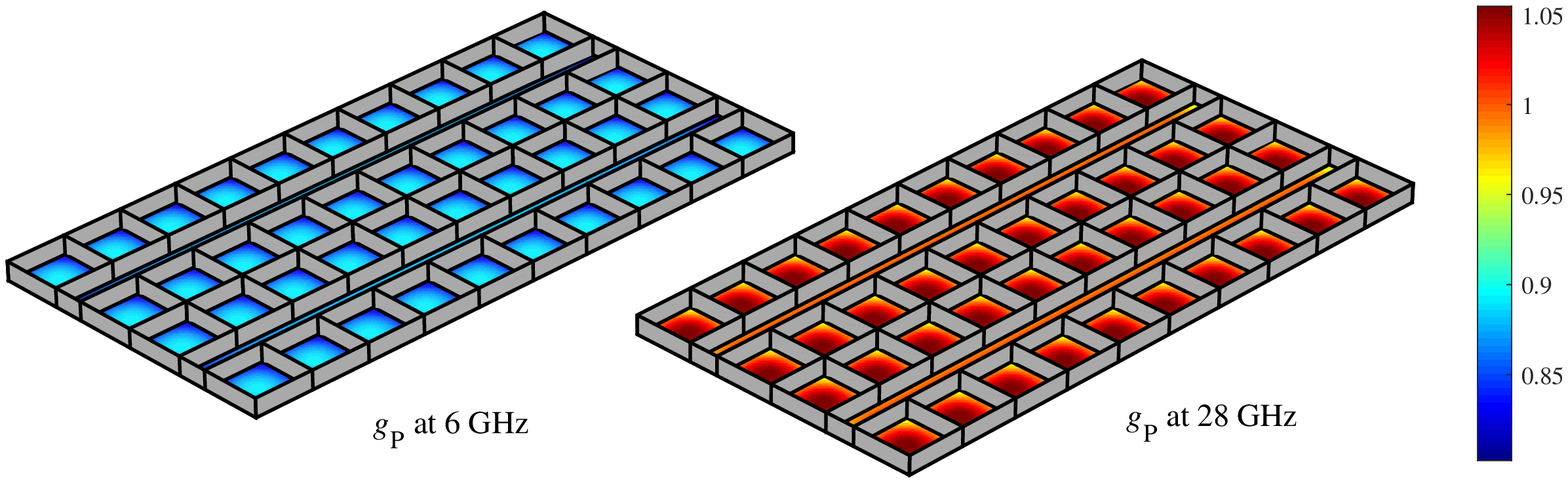}}
	\subfigure[]{\includegraphics[width=1.6in,trim=0 
		0 0 20,clip]{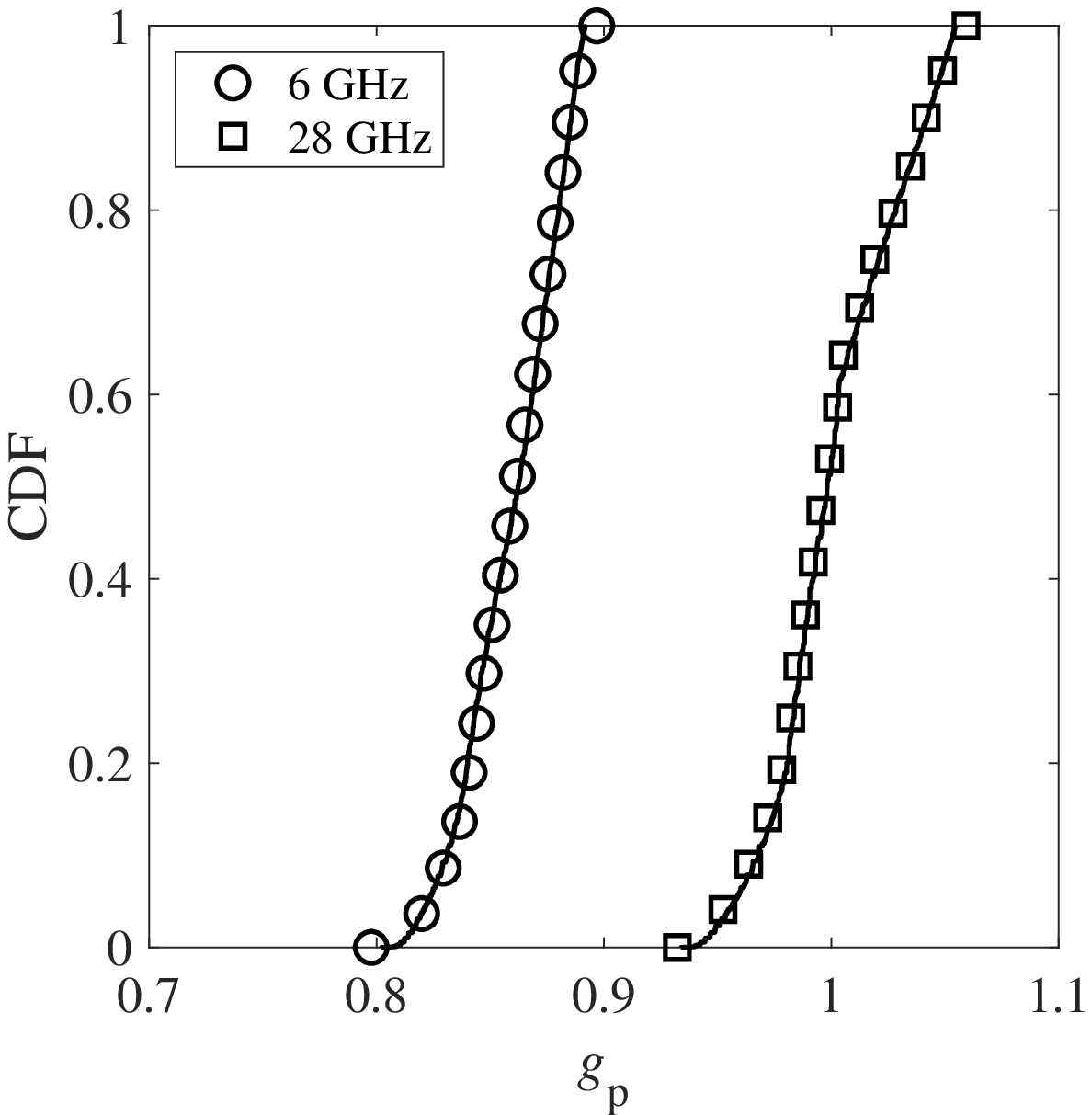}}\\
	\caption{Evaluation of $g_{\mathrm{I}}$ and $g_{\mathrm{P}}$ in the WINNER~II A1 building. 
(a) Distributions of $g_{\mathrm{I}}$. 
(b) CDFs of $g_{\mathrm{I}}$, where the lines are computed by the analytical models and the markers are obtained via the machine-learning-based empirical model.
(c) Distributions of $g_{\mathrm{P}}$. 
(d) CDFs of $g_{\mathrm{P}}$, where the lines are computed by the analytical models and the markers are obtained via the machine-learning-based empirical model.
}
	\label{A1analysis100}
\end{figure*}

\subsection{Numerical results and discussion}

\subsubsection{BWP in single rooms}
In Fig.~\ref{gigp6G}(a-d), the evaluations of single rectangular rooms with different dimensions at 28~GHz and 6~GHz are illustrated.
For the sake of comparison, average $g_{\mathrm{I}}$ and $g_{\mathrm{P}}$ computed over uniformly distributed UE locations in the rooms are given in Fig.~\ref{gigp6G}(e-f). 

From Fig.~\ref{gigp6G}(a-b), where $R_{\mathrm{O}}=5.39$ m, $R_{\mathrm{L}}=7.01$ m, and $R_{\mathrm{N}}=2.88$ m at 28~GHz, we make the following observations:
\begin{itemize}
	\item For a large room, e.g., 60-100 $\mathrm{m^2}$, with a small aspect ratio (AR), e.g., 1 or 2, if the reference UE is located in the center of the room, most LOS signals are received as intended signals. 
	Whereas if the reference UE is located close to the edges, more LOS interference signals from the opposite directions inside the room are received as the size of the room increases, and therefore results in a dramatically lower $g_{\mathrm{I}}$.
	
	\item For a small room, e.g., 20 $\mathrm{m^2}$, with a small AR, e.g., 1 or 2, $R_{\mathrm{L}}$ is larger than the maximum dimension of the room, and therefore $I_{\mathrm{L}}=0$. 
	Then, if the reference UE is located in the center of the room, NLOS interference signals can be received from all directions. 
	Whereas if the reference UE is close to edges, NLOS interference signals from the opposite directions can be blocked.
	Therefore, $g_{\mathrm{I}}$ at the room center is smaller than closer to the edges.
	
	\item For a large room, e.g., 60-100 $\mathrm{m^2}$, with a large AR, e.g., 7 or 8, the highest/lowest $g_{\mathrm{I}}$ does not appear in the center. 
	The reference UE with the highest $g_{\mathrm{I}}$ is located at a distance to the short edge, where the blockage effects on LOS interference signals are maximized.
	
	\item For rooms with small sizes, e.g., 20-40 $\mathrm{m^2}$, or large ARs, e.g., 7 or 8, more intended signals are likely to be blocked, especially for the reference UE close to edges. Thus, $g_{\mathrm{P}}$ increases with an increasing room size or a decreasing AR.
\end{itemize}
From Fig.~\ref{gigp6G}(c-d), where $R_{\mathrm{O}}=25.16$ m, $R_{\mathrm{L}}=41.63$ m, and $R_{\mathrm{N}}=7.56$ m at 6~GHz, we make the following observations:
\begin{itemize}
	\item In the 6-GHz low frequency band, $R_{\mathrm{L}}$ is larger than the maximum dimension of all sample rooms, i.e., $I_{\mathrm{L}}=0$, which means that no strong LOS interference will reduce $g_{\mathrm{I}}$. 
	Therefore, $g_{\mathrm{I}}$ varies within the small interval between 41 and 46. 
	We conclude that the room properties, i.e., size and AR, have negligible impact on $g_{\mathrm{I}}$ in this situation. 
	
	\item Compared with the 28-GHz band, the 6-GHz low frequency band has a lower  $g_{\mathrm{P}}$, since more LOS intended signals are blocked by the building structures.
	
	\item As at the 28-GHz band, more intended signals are likely to be blocked by building structures for rooms with small sizes or large ARs. 
	Therefore, $g_{\mathrm{P}}$ increases with an increasing room size or a decreasing AR. 
\end{itemize}

Fig.~\ref{gigp6G}(g-h) show the average $g_{\mathrm{I}}$ and $g_{\mathrm{P}}$ as a function of the operating frequency.
$g_{\mathrm{P}}$ increases with an increasing operating frequency because the intended signals are less likely to be blocked due to a smaller coverage distance in higher frequency bands. 
Whereas, $g_{\mathrm{I}}$ first increases and then decreases, as more NLOS interference signals can be blocked yet more LOS interference signals, which are much stronger, will be received by UE. 
Therefore, an optimum $g_{\mathrm{I}}$ exists regarding the operating frequency, which can be analytically derived by the algorithm proposed in \cite{IG}.

\subsubsection{BWP of a building given its layout}
The BWP evaluation of a building is demonstrated in a typical office building assuming A1 scenario of WINNER~II channel model \cite[Fig. 2.1]{WINNER}. 
$g_{\mathrm{I}}$ and $g_{\mathrm{P}}$ for each room and corridor obtained by the analytical models proposed in \cite{BWP} are illustrated in Fig.~\ref{A1analysis100}(a) and Fig.~\ref{A1analysis100}(c), respectively. 
The cumulative distribution functions (CDFs) of $g_{\mathrm{I}}$ and $g_{\mathrm{P}}$ are plotted in Fig.~\ref{A1analysis100}(b) and Fig.~\ref{A1analysis100}(d), respectively, where the markers are obtained by the trained ANNs and show an accurate prediction. The root-mean-square errors of predicting $g_{\mathrm{I}}$ and $g_{\mathrm{P}}$ are 0.32 and 0.002 at 6 GHz, and 0.64 and 0.003 at 28 GHz, respectively.
At 6~GHz, the average $ g_{\mathrm{I}}=41.74$ and the average $g_{\mathrm{P}}= 0.86$. 
At 28~GHz, the average $g_{\mathrm{I}}=24.33$ and the average $g_{\mathrm{P}}= 1.001$.
The assumed building at 6~GHz significantly outperforms it at 28~GHz, i.e., in the mm-Wave band, in terms of $g_{\mathrm{I}}$ since it isolates the reference UE from almost all the interference signals. 
Whereas, $g_{\mathrm{P}}$ in the two frequency bands shows a limited difference because the LOS intended signal power from transmit elements around the reference UE, which takes a dominant portion of the intended signal power, is barely blocked in a building assuming the WINNER II A1 scenario at both 6~GHz and 28~GHz. 
The numerical results have shown that the variations of $g_{\mathrm{I}}$ and $g_{\mathrm{P}}$ properly capture the effects of building structures on the network performance. 
In \cite{multiwall}, we use the multi-wall path gain model to compute $g_\mathrm{I}$ and $g_\mathrm{P}$, where the NLOS links are further distinguished by the exact number of intersections \cite{MWM}. 
This multi-wall-model-based BWP evaluation scheme can be employed to address the impact of the relative location of a room in the building, especially in lower frequency bands.

\begin{figure*}[t]
	\centering
    \includegraphics[width= 6.2 in,trim=0 0 0 0,clip]{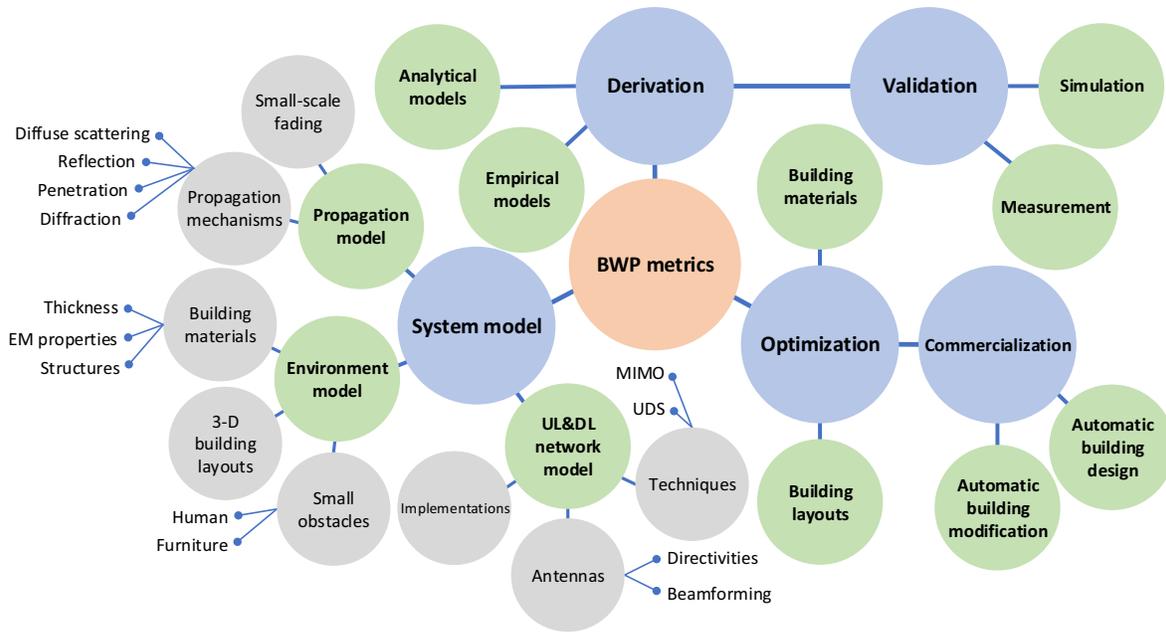}
	\caption{Outline of future working directions.} 
	\label{Future}
\end{figure*}

\section{Open issues}

The numerical results have shown the potential of the novel concept. 
Nevertheless, the idealistic network model and simplified building structure descriptions may pose certain practical challenges for the BWP evaluation scheme. 
In this section, we propose possible future working directions to complement the BWP framework in terms of  the metrics, system model, derivation schemes, and validation approaches. 
Furthermore, we discuss the potential research directions in BWP optimization and introduce our vision in future automatic wireless friendly building design.
The schematic representation of possible future working directions is outlined in Fig.~\ref{Future}.

\subsection{Design new BWP metrics}

Network performance metrics, beyond the SINR, such as throughput and wireless connection latency, are indeed relevant at the design stage of a building. 
Therefore, new BWP metrics need to be defined appropriately. 
For example, two new BWP metrics could be referred to as the \textit{throughput gain} and the \textit{latency gain}, defined as the effective increase of the throughput and the effective reduction of the connection latency caused by the presence of a building in comparison to the open-space scenario, respectively. 
The relationships among the various BWP metrics will be investigated, and a more mature BWP evaluation framework considering all the essential metrics will be developed.

\subsection{System model}

A more comprehensive system model should be employed to enable the BWP metrics to capture the impact of overall building structures on practical in-building wireless networks. 
For the practical network model, it is significant to go beyond the idealistic assumptions in previous works. 
For example, future works ought to consider the intrinsic constraints of different techniques in both downlink and uplink transmissions, such as the density/scale for the UDS network/MIMO system, implementations, antenna directivities and beamforming techniques. 
Naturally, in addition to the ARs and sizes of rooms addressed in previous works, more factors of the building structures affecting indoor networks can also be considered in the environment model; for example, BWP metrics need to be investigated for 3-D building layouts with consideration of inter-floor connections/interference.
The impact of the properties of building materials on the BWP metrics is another crucial open issue to be investigated in the future. Moreover, the effects of indoor small objects should be considered, such as furniture and humans, especially in the mm-Wave bands.
The propagation model should be defined corresponding to the network and environment models, where the small-scale fading should be included if not negligible.
As the BWP framework has to guide building design without knowing specific wireless techniques and device configurations, it is challenging to define a proper network model using general parameters available at the building design stage. 
Devising future-proof frameworks should therefore be the ultimate goal.   

\subsection{Derivation and validation}

A more complicated system model will make it more difficult to derive and maintain the tractability of the analytical models of the BWP metrics.
To lower the computation burden, empirical models can be developed to predict the BWP. 
The importance of the properties of building structures needs to be quantitatively evaluated to extract a finite set of input parameters for the BWP prediction. 
One feasible form of the empirical model is to map the input parameters to the BWP metrics by a mathematical function and obtain its coefficients by curve fitting. 
It can simplify the computation significantly, thereby helping the BWP evaluation to be widely implemented.  
Furthermore, the promising machine-learning-based approaches can be employed to develop empirical models with high prediction accuracy.
Other than the dimensions of the rectangular rooms used in the ANNs introduced in Section~III, additional input parameters could be introduced to improve the accuracy and generality of the machine-learning-based models.

The effectiveness of the BWP evaluation schemes should be verified in different scenarios. 
In our previous works, the proposed analytical models have been validated by the Monte Carlo simulations, where the assumed idealistic network model had been represented by a large number of randomly placed transmit elements \cite{BWP, IG,multiwall}. 
In future works, the ray-based deterministic simulations following the geometrical optics theory can be employed to verify the proposed BWP evaluation framework under more practical network and environment models. It is also necessary to validate the proposed methods with practical and real-life experiments. Therefore, the BWP research offers new exciting challenges to properly design new wireless friendly buildings with experimentally verifiable practical performance.

\subsection{Optimization} 

The material selections and the layout design of a building are subject to the constraints from other essential building utilities, such as safety, insulation, and daylighting, leading to enormous complexity of optimizing building structures to improve the BWP. 
Here we discuss three potential research directions in BWP optimization.

\subsubsection{Optimization of building materials}
Changing the materials of non-loadbearing partition walls can be considered one of important solutions to modify the building design for better BWP. 
This has a minor impact on the function of a building compared to changing the building layout. Furthermore, advanced manufacture and construction practices, such as 3-D concrete printing, metamaterials, etc., may provide multiple new building material design strategies to satisfy various specific requirements. The optimization of material selections to partition walls may significantly improve the BWP with a limited cost. However, once the optimal material selection to partition walls cannot make the building achieve adequate BWP, the building layouts and building materials still need to be considered simultaneously.

\subsubsection{Joint optimization of network configurations and building structures}

Modular construction is one of the promising techniques to fulfill the increasing housing demand.  
It facilitates the off-site manufacture of the modules complete with essential services and internal finishes for walls, floors, and ceilings, thereby significantly improving the construction productivity, quality, and efficiency.  
We foresee that the wireless transmit elements are expected to be pre-installed on or even embedded in building materials in the prefabricated modules to make up the wireless system in a building. 
Thus the building structures and the wireless network should be designed simultaneously for a modular construction building.
In future works, instead of searching optimal network configurations for existing buildings \cite{IG}, a dedicated network solution maximizing the BWP should be determined while designing the building.  

\subsubsection{Joint optimization of buildings in a neighborhood}
In dense urban scenarios, an in-building wireless network can be seen as a part of the wireless system consisting of indoor and outdoor transmit elements in the neighborhood where the building is located. 
Indoor users could be served by outdoor BSs, while indoor transmit elements could also serve outdoor users or cause leakage.
The BWP of buildings in a neighborhood could be jointly optimized.  
Instead of considering an individual building, the joint optimization of a neighborhood could further exploit the signal power from outdoor BSs and the blockage effects of surrounding buildings. 
Correspondingly, the complexity of the optimization could rise significantly. 

\subsection{Application and commercialization}

In future wireless friendly building design, reference levels of BWP metrics will be defined for buildings differing in types and dimensions. 
The BWP evaluation and optimization process will be iteratively conducted in building design to search for a solution satisfying the requirements from both construction and wireless communication perspectives. 
To enable the BWP framework to be widely employed in the construction industry, our vision is to develop a complete BWP evaluation and optimization system and integrate it into an automatic building configurator to automatically generate or modify the building structures with respect to the targeted values of BWP metrics. 
A full-flared software tool could create massive academic and commercial value, but it is challenging to meet all the needs of architects and civil engineers. 

\section{Conclusions}

This article discusses the novel concept of building wireless performance or BWP in short. 
The BWP has been introduced to quantify the effects of building structures on indoor wireless network performance. 
We have defined two BWP metrics labeled as the interference gain and the power gain, or the IG and the PG, to quantify the impact of building structures on the power levels of interference and intended signals, respectively.
The proposed framework has been employed to explore the impact of the rooms' sizes and aspect ratios on the BWP. 
We have discussed future works in this research direction in terms of the new metrics, system model, derivation schemes, and validation approaches. 
Moreover, we have introduced potential research directions in BWP optimization and explained our vision in future automatic wireless friendly building design towards 6G and beyond indoor wireless communications.

{\small
\textbf{Jiliang Zhang} [M'15, SM’19] received the B.E., M.E., and Ph.D. degrees from the Harbin Institute of Technology, Harbin, China, in 2007, 2009, and 2014, respectively. He is now a KTP Associate at the Department of Electronic and Electrical Engineering, The University of Sheffield, Sheffield, UK. His current research interests include, but are not limited to wireless channel modelling, modulation system, relay system, vehicular communications, ultra-dense small cell networks, and smart environment modelling.

\textbf{Andr\'{e}s Alay\'{o}n Glazunov} [SM’11] received the M.Sc., Ph.D., and Docent qualifications, in 1994, 2009, and 2017, respectively. He is an Associate Professor with the Department of Electrical Engineering, University of Twente, Enschede, Netherlands, and an Affiliate Associate Professor with the Chalmers University of Technology, Gothenburg, Sweden. Dr. Glazunov is one of the pioneers producing the first standardized spatial channel models and OTA measurement techniques for 3GPP, and devising novel OTA techniques, e.g., the random-LOS and the hybrid antenna characterization techniques. His current research interest is focused on the design and characterization of smart antenna systems and smart wireless propagation environments.

\textbf{Wenfei Yang} received the B.E. degree in 2016 from Harbin Institute of Technology, Weihai, China, and the M.E. degree in 2017 from the University of Sheffield, Sheffield, UK, where she is currently pursuing the Ph.D. degree with the Department of Electronic and Electrical Engineering. Her current research interests include wireless channel modeling, ultra-dense small cell networks, and smart environment modeling. 

\textbf{Jie Zhang} has held the Chair in Wireless Systems at the University of Sheffield since 2011. He is Founder, Board Director and Chief Scientific Officer of Ranplan Wireless, a company listed on Nasdaq OMX. Along with his students and colleagues, he has pioneered research in small cell and heterogeneous network (HetNet) and published some of the landmark papers and books on these topics. Since 2010, he and his team have developed ground-breaking work in modelling and designing smart built environments considering both wireless and energy efficiency. His Google scholar citations are in excess of 7900 with an H-index of 37.

\ }
\end{document}